\documentclass[aps,prc,twocolumn,showpacs,nofootinbib]{revtex4}%
\usepackage{amsfonts}
\usepackage{amsmath}
\usepackage{amssymb}
\usepackage{epsfig}
\usepackage{graphicx}
\usepackage{dcolumn}%
\def\bea{\begin{eqnarray}}
\def\eea{\end{eqnarray}}

\begin{document}
\title{Glueball enhancements in  {\mbox {\boldmath${p(\gamma,
VV)p}$\unboldmath}}
through vector meson dominance}

\author{Stephen R. Cotanch}
\affiliation{Department of Physics, North Carolina State University,
Raleigh, NC 27695-8202}
\author{Robert A. Williams}
\affiliation{ Nuclear Physics 
Group, Hampton University,
Hampton, VA 23668 \\
and
Jefferson Lab,
12000 Jefferson Ave, Newport News, VA 23606 }
%\pacs{ }
\begin{abstract}
Double vector meson photoproduction, $p(\gamma,G \rightarrow
V V)p$,
mediated by a scalar glueball  $G$ is investigated. Using vector meson
dominance (VMD)  and Regge/pomeron phenomenology,  a measureable
glueball  enhancement is predicted in the invariant $VV = \rho \rho$ and $\omega \omega$ mass
spectra.  The scalar glueball is assumed to be the lightest physical state
on the daughter  pomeron trajectory governing  diffractive  vector meson
photoproduction.  In addition to cross sections, calculations for  hadronic and
electromagnetic glueball decays,
$G \rightarrow V V'$ ($V, V' = \rho, \omega, \phi,\gamma$),  and 
$\gamma_v V \rightarrow G$ transition form factors are presented based upon
flavor universality, VMD and phenomenological couplings from  
$\phi$ photoproduction analyses.
The predicted glueball decay widths are similar to an independent
theoretical study. A novel signature for glueball detection is
also discussed. 
\end{abstract}
\volumeyear{year}
\volumenumber{number}
\issuenumber{number}
\eid{identifier}
\date{\today}
\startpage{1}
\endpage{2}
\maketitle

%\preprint{HEP/123-qed}

%\date{\today}

%\date[Date text]{date}
%\received[Received text]{date}

%\revised[Revised text]{date}

%\accepted[Accepted text]{date}

%\published[Published text]{date}

\section{Introduction}

Even though Quantum Chromodynamics (QCD) is the accepted theory of hadronic
physics,
realistic nonperturbative QCD predictions for reaction amplitudes are still not
available. However,  Quantum Hadrodynamic  (QHD) calculations continue to
provide
a reasonable framework for the analysis of data.  Related, the historical success of
vector meson dominance (VMD) and Regge theory has led to an established legacy for 
investigating
both electromagnetic and hadronic processes. Because of the wide interest in the gluonic aspects
of QCD, especially glueballs, and new experimental opportunities at electromagnetic accelerator
facilities, such as Jefferson Lab, this work
combines QHD, VMD and Regge/pomeron physics to study double vector meson photoproduction,
$p(\gamma,G \rightarrow V V)p$, mediated by a scalar glueball,
$G$.

Central to our formulation is the pomeron-glueball hypothesis
(PGH)~\cite{flsc,erice,scfb17}  which connects the pomeron with the even signature  
$J^{PC} =$ $2^{++}, 4^{++} ...$ glueball Regge trajectory. 
Indeed both
theoretical~\cite{flsc,erice,scfb17,kiss,liu,meyer} and
experimental~\cite{koc} evidence continues to accumulate which supports
this conjecture.   The PGH provides an attractive, logical  framework for determining all
glueball-hadron couplings from  established  pomeron phenomenology.

In this study we use the PGH to
extend  the effective Lagrangian model developed  for
$\phi$ photoproduction~\cite{RAWphi} and time-like virtual Compton scattering
(TVCS)~\cite{TVCS}, to double vector meson photoproduction mediated by a
scalar glueball.
The necessary glueball-vector meson ($V = \rho, \omega, \phi$) hadronic and
electromagnetic couplings
are uniquely determined from PGH, VMD and isospin
symmetry (flavor independence) of the glueball-hadron couplings.  
In addition to cross sections,
we predict the $J^{PC} = 0^{++}$ glueball partial decay widths for
double vector,
$G \rightarrow V V'$, one photon, $G
\rightarrow V \gamma$,  and two photon, $G \rightarrow \gamma \gamma$,  decay
channels. Since  the vector meson leptonic decay constants are known, we
also apply
VMD to derive the radiative,  $\gamma_v V \rightarrow G$, transition form
factors required for scalar glueball electroproduction
calculations.
Our key finding is the prediction of a measurable $p(\gamma,G \rightarrow V V)p$   cross section
and a glueball enhancement in the 
$\rho \rho$ and $\omega \omega$  invariant mass spectra 
near 1.7 GeV. 

This paper spans six sections. In section II we review the essential
features of $\phi$ electromagnetic production and TVCS~\cite{TVCS} that
are relevant for formulating
$VV$ photoproduction.  Then we detail the QHD model in section III
and in section IV present the VMD relations,   glueball radiative
transition form factors and decay widths.  Section V contains our main results
with theoretical $p(\gamma,VV)p$  cross sections documenting a measurable
glueball enhancement  and novel signature decay.  Finally,  in section VI we  
summarize and comment
on future investigations.

%\newpage
\section{$\phi$ Photoproduction and TVCS Model Summary}

Vector meson photoproduction is known to be dominated by diffractive
scattering
at low momentum transfer and high energy.  The diffractive amplitude has a
clear
exponential $t$-dependence, presumably generated by a tower of gluon
$t$-channel exchanges, collectively known as the pomeron~\cite{Collins}.
At low energy and for large momentum transfer, vector meson photoproduction
is complicated
by non-diffractive mechanisms such as pseudoscalar meson ($\pi, \eta, \eta'$)
exchange~\cite{RAWphi,Pichowsky},
nucleon resonances and
two-gluon exchange~\cite{Laget}.

In $\phi$ photoproduction there are additional non-diffractive amplitudes
due to
strangeness knockout~\cite{Henley,Titov} and Okubo-Zweig-Iizuka
(OZI)~\cite{OZI}
violating/evading $\phi N$ couplings~\cite{RAWphi}. The
$\phi$ photoproduction reaction
is especially interesting for probing the intrinsic strangeness content of the
nucleon~\cite{Henley,Titov}
and yields important constraints for   the
(off-shell) nucleon form factors in the vector meson resonance region
accessible in
TVCS, $\gamma p \rightarrow e^+ e^- p$~\cite{TVCS}.
For example, OZI evading $\phi N$ vector and tensor couplings
contribute to the nucleon strangeness radius, strange magnetic moment
and  provide an improved description
of the neutron electric form factor,
$G_{E}^{n}(q^2)$~\cite{Hohler,Dubnicka,Gari,WKL,WPT}.

Our previous results documented that precision
$\phi$ photoproduction and di-lepton TVCS data near the $\phi$ production
threshold
will provide important constraints for disentangling the complicated
diffractive/non-diffractive amplitude components. In this work we
apply  the same effective Lagrangian used to calculate $t$-channel
pomeron exchange in $p(\gamma,V)p$, to  scalar glueball
photoproduction
$p(\gamma,G \rightarrow V V)p$. An important feature in the
photoproduction/TVCS model is  the photon-pomeron-vector meson vertex
coupling associated with the $t$-channel pomeron exchange.
Again we utilize this and now interchange the role of the pomeron and
vector meson to consider $t$-channel $\rho$, $\omega$ and $\phi$
exchange  leading to pomeron, or via the PGH, glueball
photoproduction.  We also interpret the scalar
glueball as the $J=0$  physical state on the daughter pomeron trajectory.

\section{QHD Model Details}

We formulate the double vector meson photoproduction
reaction
as a two step mechanism mediated by a  scalar glueball, $G$
\begin{eqnarray}
\gamma(q,\lambda) + p(p,\sigma) & \longrightarrow & p(p',\sigma') +
G(q')  \nonumber \\
 & \longrightarrow &  p(p',\sigma') + V(v_1,\lambda_1) + V'(v_2,\lambda_2)   \nonumber
\end{eqnarray}
where the energy-momentum 4-vectors (helicities) for the photon, 
proton, glueball,  recoil proton and vector mesons
are given by $q$ ($\lambda$), $p$ ($\sigma$), $q' = v_1 + v_2$,  $p'$ ($\sigma'$)
and $v_{i = 1, 2}$ ($\lambda_i)$, respectively.
The  general case is considered involving photoproduction of a
glueball that may be on, $M_G = \sqrt{{q'}^2}$, or off-shell (virtual), 
$M_G  \ne \sqrt{{q'}^2}$, and
decays into two, possibly different, vector mesons
$VV'= \rho \rho$, $\omega \omega$, $\phi \phi$ or $\omega \phi$ having
masses $M_V = \sqrt{v_1^2}$, $M_{V'} =
\sqrt{v_2^2}$. The 3-body final state differential cross section factorizes
\begin{equation}
\frac{d\sigma}{dt \; dM_{VV'}} \;= \; \frac{d\sigma_v}{dt} \; {\cal F}_{<VV'|G>}
\end{equation}
where $M_{VV'} = \sqrt{{q'}^2}$ is the  invariant $VV'$ mass and $\frac{d\sigma_v}{dt}$ is the  
virtual glueball photoproduction cross section
\begin{equation}
\frac{d\sigma_v}{dt} \;=\; \frac{\pi}{\omega_{cm}^2} \;
|<G \;p\;| \;\hat{T}\; |\;\gamma \; p>|^2 \ .
\end{equation}
The vector meson flux, ${\cal F}_{<VV'|G>}$,    resulting from the
glueball decay can be expressed in terms of phase space, ${\cal P}_{VV'}$, the glueball
propagator,
$\Pi_G(q')$, and the $G \rightarrow V V'$ decay amplitude, $<V V'|G>$,
\begin{eqnarray}
{\cal F}_{<VV'|G>} &=&  {\cal P}_{VV'} \; |\;\Pi_G(q')\;|^2 \; |<V V'|G>|^2 \\
{\cal P}_{VV'} &=&  \frac{|{\bf q}'|}{256 \pi^4 M_p^2} \;
%\left(\frac{\omega_{cm}}{\omega_{lab}}\right)^2
\frac{\omega^2_{cm}}{\omega^2_{lab}} \\
\Pi_G({q'}^2) &=& \frac{\sqrt{{q'}^2}}{{q'}^2 - M_G^2 + i \sqrt{{q'}^2}\; \Gamma_G} \;
\left(\frac{s-s_{th}}{s_{0}}\right)^{\alpha_{\cal P}({q'}^2)} \label{pomprop}
\\ <V V'|G> &=& \frac{g_{GVV'}}{2 M_0} \; F^V_{\mu \nu}(v_1,\lambda_1) \;
F_{V'}^{\mu \nu}(v_2,\lambda_2) \ .
\end{eqnarray}
In these equations $\omega$ refers to the photon energy (in the appropriate
frame),
$M_p$ is the proton mass, $\Gamma_G$ is the glueball total
width,
$g_{GVV'}$ is the glueball-vector meson  coupling constant,
$M_0$ is a reference mass (set to 1 GeV) which permits
a dimensionless glueball coupling and $F_{\mu \nu}^V$
is the vector meson current tensor specified below.
The effective  glueball propagator
is a generalization of the empiracle space-like pomeron 
prescription~\cite{RAWphi}
%\begin{equation}
%\Pi_{{\cal P}}(t) \;=\; \frac{e^{\beta t}}{t - M_{{\cal P}}^2} \; .
%\label{e1pP}
%\end{equation}
with the
pole mass fixed at $M_G = 1.7 $ GeV,
consistent with the generally accepted lightest scalar glueball.
%and modified by an empirical exponential factor,
%$e^{\beta t}$, which reproduces the established
%diffractive photoproduction $t$-dependence.
Following Ref.~\cite{Collins},    we have included in
Eq. (\ref{pomprop}) the Regge factor,
$(\frac{s-s_{th}}{s_{0}})^{\alpha_{\cal P}({q'}^2)}$,  which
describes the high energy behavior.
Here  $s =
(q+p)^2$ is the usual $cm$ energy Mandelstam variable and
$\alpha_{\cal P}({q'}^2)$ is the pomeron trajectory of even signature glueballs
with established
linear form $\alpha_{\cal P}(t) = \alpha_{0} + \alpha' t$.
Because Regge theory only governs the asymptotic high energy
behavior,  we introduce
the  parameter, $s_{th}$ ($0 \leq s_{th} \leq s_{0}$), to describe the low energy
double meson production amplitude  
with the reference energy, $\sqrt{s_{0}}$, fixed at the threshold, 
$s_{0} = (M_p + M_{V} + M_{V'})^2$. In  previous, successful analyses of 
$\phi$ photoproduction
using this  prescription~\cite{RAWphi,TVCS}
the available data clearly selected the maximum value, $s_{th} = s_{0}$,
which is used through out this paper.  If we omit the Regge factor the effective gluonic
propagator takes a standard hadron (glueball) form and thus we loosely distinguish between
pomeron (Regge) mediated or glueball (non-Regge) production.  In section V we compare cross
section predictions for both propagators.

In the helicity representation the glueball photoproduction  amplitude,  $<G \;p\;|
\;\hat{T}\; |\;\gamma
\; p>$, is  
\begin{eqnarray}
<G \;p\;| \;\hat{T}\; |\;\gamma \; p> &=&
 \; \epsilon_{\mu}(\lambda) \;
{\cal H}^{\mu}_{\sigma' \sigma} \; ,
\end{eqnarray}
where $\epsilon (\lambda)$
is the  photon
polarization 4-vector in the helicity basis and
${\cal H}^{\mu}_{\sigma' \sigma}$ is the hadronic current
obtained by application of Feynman rules to the tree level $s$, 
$t = (q' - q)^2$ and
$u = (p' - q)^2$ channel QHD diagrams. The hadronic current is evaluated
in the total $cm$ system (${\bf q + p = q^{\prime} + p'} = 0$)
with the z-axis taken along ${\bf q}$.
In this frame the two photon polarization vectors are
\begin{eqnarray}
\epsilon (\lambda) &=&
 - \frac{\lambda}{\sqrt{2}} (0,1,i\lambda,0)  \;\;\;\;\;\;
 (\lambda = \pm) \ .
\end{eqnarray}
The $G(0^{++})\rightarrow V(1^{--}) V'(1^{--})$ decay helicity amplitude,
$<VV'|G>$, involves the vector
meson current tensors $F^V_{\mu \nu}(v_1,\lambda_1)$ and
$F_{V'}^{\mu \nu}(v_2,\lambda_2)$ given by 
\begin{eqnarray}
F^V_{\mu \nu}(v_i,\lambda_i) &=& v_{i\mu} \epsilon_{i \nu}(v_i,\lambda_i)
\;-\;
v_{i\nu} \epsilon_{i \mu}(v_i,\lambda_i)   
\end{eqnarray}
with  spin polarization 4-vectors, $\epsilon_{i} (v_i,
\lambda_i)$, subject to the Lorentz condition $v_i \cdot \epsilon_i = 0$ for $i = 1,2$.
These polarization vectors satisfy 
\begin{eqnarray}
\sum_{\lambda_i} 
\epsilon_i^{\mu}(v_i, \lambda_i) {\epsilon_i^{\nu}}^* (v_i, \lambda_i) = - g^{\mu \nu} +
v_i^{\mu} v_i^{\nu}/M_i^2
\label{metric}
\end{eqnarray}
where $g_{\mu \nu} = g^{\mu \nu}$ is the standard metric tensor.
The invariant helicity decay amplitude involves the contraction 
\begin{eqnarray}
%F^V(v_1,\lambda_1) &\cdot&  F_{V'}(v_2,\lambda_2) \;=\; \\
%&2&  \left (v_1 \cdot v_2 \; \epsilon_1 \cdot \epsilon_2  
% \; - \;  v_1 \cdot \epsilon_2 \; v_2 \cdot \epsilon_1 \right ) \nonumber .
F^V_{\mu \nu}F_{V'}^{\mu \nu} =
2  \left (v_1 \cdot v_2 \; \epsilon_1 \cdot \epsilon_2  
\; - \;  v_1 \cdot \epsilon_2 \; v_2 \cdot \epsilon_1 \right )  
%\nonumber .
\end{eqnarray}
and can be evaluated in the glueball rest frame where,
with ${\bf v}_1$  along the z-axis, the
mesons 4-momenta are
\begin{eqnarray}
v_1 &=& (E_1, {\bf v}_1) = (E_1,0,0,k_V) \\
v_2 &=& (E_2, {\bf v}_2) = (E_2,0,0,-k_V) \ .
\end{eqnarray}
The 3-momentum $k_V = |{\bf v}_1| = |{\bf v}_2|$ depends on
the vector meson  and the  glueball (invariant $VV'$) masses
%\begin{equation}
\begin{eqnarray}
k_V = \frac{1}{2 \sqrt{{q'}^2}} 
\left[ ({q'}^2 + M_V^2 - M_{V'}^2)^2 - 4 {q'}^2 M_V^2 \right]^{\frac{1}{2}} . 
\label{kv} 
%\end{equation}
%\begin{equation}
% &=& \frac{1}{2 } \;
%\left[ {q'}^2 - 4  M_V^2 \right]^{\frac{1}{2}}   \  (M_V = M_{V'}) \ .
%\end{equation}
\end{eqnarray}
In this frame the meson polarization vectors are 
\begin{eqnarray}
\epsilon_1(\lambda_1 = \pm) &=& - \frac{\lambda_1}{\sqrt{2}} 
(0,1,i\lambda_1,0) \\
\epsilon_1(\lambda_1 = 0) &=& \frac{1}{M_V} 
(k_V,0,0,E_1) \\
\epsilon_2(\lambda_2=\pm) &=& - \frac{\lambda_2}{
\sqrt{2}} (0,1,i \lambda_2,0) \\
\epsilon_2(\lambda_2 = 0) &=& \frac{1}{M_{V'}}
(-k_V,0,0,E_2) \ .
\end{eqnarray}
If the spins of the final state  mesons are not detected, the cross section entails
a  helicity sum giving the factor
\begin{eqnarray} 
\lefteqn{ S \equiv \sum_{\lambda_1 \lambda_2 = 0, \pm 1}  |<VV'|G>|^2 } \\
 &&    = \frac{g^2_{GVV'}}{M^2_0}  
\sum_{\lambda_1
\lambda_2 = 0, \pm 1} \;
\left[\; v_1 \cdot v_2 \; \epsilon_1(\lambda_1) \cdot \epsilon_2(\lambda_2) \right. \nonumber \\
 &&  \left.  \;-\;
v_1 \cdot \epsilon_2(\lambda_2) \; v_2 \cdot \epsilon_1(\lambda_1)
\;\right]^2 \ .
\end{eqnarray}
Using the above specific kinematical representation for $v_i$ and $\epsilon_i$ or,
more generally for any frame, Eq. (\ref{metric}), the summation reduces to
the invariant result
\begin{eqnarray} 
S &=& \frac{g^2_{GVV'}}{M^2_0} \left[2 (v_1
\cdot v_2)^2 + M^2_V M^2_{V'} \right] \\
&\rightarrow& \frac{g^2_{GVV}}{2M^2_0} \left[ (M_{VV}^2 - 2 M^2_V)^2 + 2 M^4_V \right] (V =
V')  .
%= 3 \; \frac{g^2_{GVV'}}{M^2_0} \;
%\left[\; k_V^2 + E_1 E_2  + \frac{k_V^2 {q'}^2}{M_V M_{V'}} \ .
%\;\right]^2 \ .
\end{eqnarray}

The effective QHD Lagrangian  for
the strong and electromagnetic interactions
generates the following contributions to the hadronic current
\\

\underline{$t$-channel $V=\rho,\omega,\phi$ exchange amplitudes:}
\begin{eqnarray}
{\cal H}^{\mu}_{\sigma' \sigma} &=& e \; g_{VNN} \;
 \frac{ \kappa_{GV\gamma}}{M_0}
 \;  F_{t}(t;\lambda_{cut}) \; 
  \nonumber \\  & & \times \; \bar{u}(p',\sigma') \; [\; \gamma^{\mu} \; + \;
i \; \frac{\kappa_V^T}{M_0} \; \sigma^{\mu \alpha} q'_{\alpha} \;]
u(p,\sigma) \;
\label{eqt}
\end{eqnarray}

\underline{$s$-channel proton-glueball coupling amplitude:}
\begin{eqnarray}
{\cal H}^{\mu}_{\sigma' \sigma} &=&
e \; g_{GNN}  \; \bar{u}(p',\sigma')
\; \frac{ (p+q)\cdot \gamma + M_{p}}{s - M_{p}^2 +
\Sigma_p(s)}
 \nonumber \\ & & \times [\; \gamma^{\mu} \; + \;
i \; \frac{\kappa_p}{2 M_p} \;  \sigma^{\mu \beta} q_{\beta} \;] \;
u(p,\sigma)
\end{eqnarray}

\underline{$u$-channel proton-glueball coupling amplitude:}
\begin{eqnarray}
{\cal H}^{\mu}_{\sigma' \sigma} &=&
e \; g_{GNN}  \; \bar{u}(p',\sigma')
\; [\; \gamma^{\mu} \;+\;
i \; \frac{\kappa_p}{2 M_p} \;
\sigma^{\mu \beta} q_{\beta} \;] \nonumber \\ & & \times \;
\frac{ (p'-q)\cdot \gamma + M_{p}}{u - M_{p}^2 +
\Sigma_p(u)} \;
u(p,\sigma) \ .
\end{eqnarray}
Here  $e = \sqrt{4\pi \alpha_e}$ and
each  hadronic current term has an effective coupling strength
involving a glueball hadronic  or electromagnetic
coupling constant.
The  factor $\kappa_V^T$ in Eq. (\ref{eqt}) is the tensor to vector coupling
constant
ratio for the $\rho$, $\omega$ or $\phi$ nucleon vertex and $\kappa_p =1.793$
is the proton anomalous magnetic moment.

In the $t$-channel  the hadronic
form factor, $F_{t}(t;\lambda_{cut}) $, incorporates the
composite
nucleon and vector meson structure
which can be calculated in constituent
quark models and is important for regulating the energy and momentum transfer
dependence in meson photoproduction~\cite{Li,Lu}. However, to preserve the
covariance and crossing properties of our model, we employ the 
phenomenological form factor~\cite{RAWphi,TVCS}
\begin{eqnarray}
F_{t}(t;\lambda_{cut}) &=& \frac{ \lambda_{cut}^4 + t_{min}^2
}{\lambda_{cut}^4 + t^2} \; ,
\end{eqnarray}
normalized to unity at
$t_{min} = t(\theta^{cm}_{\gamma G} = 0)$.
Fitting the $p(\gamma,\phi)p$ data yields the optimum cut-off parameter
$\lambda_{cut} = 0.7$ GeV.
Also note that, unlike vector meson electromagnetic production, for 
$t$-channel $J^{PC} = J^{++}$
glueball (pomeron) production, pseudoscalar meson exchange is prohibited by $C$-parity
conservation.  Hence $\pi$ exchange  only contributes to the production of
$C$ = odd glueball
states which have much higher masses and are also  more difficult to detect. 

In the $s$ and $u$-channels, the highly virtual proton propagation
requires an off-shell form factor prescription. Because the $s$ and
$u$-channel
diagrams must combine to produce a conserved hadronic current, we incorporate
the off-shell effect  as a self-energy correction.
Constrained by gauge invariance and the  proton mass,
the  self-energy function must vanish at the proton pole and also be an
odd function of $s - M_p^2$. Hence we take
\begin{equation}
\Sigma_p(s) \;=\; \alpha_{off} \frac{(s - M_p^2)^3}{M_p^4}\; .
\end{equation}
The dimensionless off-shell parameter, $\alpha_{off}$ = 1.29,
was determined by fitting recent $\phi$ photoproduction data~\cite{hi-t-phi}.

The model parameters for the pomeron amplitude are listed
in Table I and the vector meson coupling constants are specified in Table II.
The glueball-vector meson couplings are assumed to be flavor
independent ({\it
universality} hypothesis) due to isospin/flavor invariance of gluonic interactions.
% Table III
%%%%%%%%%%%%%%%%%%%%%%%%%%%%%%%%%%%%%%%%%%%%%%%%%%%%%%%%%%%%%%%%%%%%%%%
%\vspace{0.15in}
\begin{table}[h]
\caption{
Pomeron/glueball hadronic coupling and Regge trajectory,
$\alpha(t) = \alpha_0 + \alpha' t$, parameters. }
\begin{center}
\begin{tabular}{|c|c|c|c|}
\hline
 $ \; g_{GNN} \;$
 &  $ \; g_{GVV'} \;$
&$ \; \alpha_{0} \; $ & $\; \alpha' \; (GeV^{-2}) \; $ \\
\hline
44.0 & 3.43  & $\; 1.0 \;$ & 0.27  \\
\hline
\end{tabular}
\end{center}
\end{table}

%\newpage
% Table IV
%%%%%%%%%%%%%%%%%%%%%%%%%%%%%%%%%%%%%%%%%%%%%%%%%%%%%%%%%%%%%%%%%%%%%%%
\begin{table}[h]
\caption{Vector meson coupling constants.}
\begin{center}
\begin{tabular}{|c|c|c|c|c|c|}
\hline
$ \; g_{\rho NN} \;$ & $ \; g_{\omega NN} \;$ & $ \;g_{\phi NN} \;$
& $\; \kappa_{\rho}^{T} \;$ &
$\; \kappa_{\omega}^{T} \;$  & $\; \kappa_{\phi}^{T} \;$  \\
\hline
2.014 & 3.411  & 1.306 & $\; 6.100 \;$ & $\; 0.140 \;$  & $\; 1.820 \;$ \\
\hline
\end{tabular}
\end{center}
\end{table}

\section{Glueball Decay widths and Transition Form Factors}

In this section we present  our VMD formulation for the proton,
vector meson and
glueball transition  form factors.
In our previous $\phi$ photoproduction/TVCS calculations we utilized a
hybrid VMD
approach~\cite{WKL,WPT} that was a generalization of the
model developed by Gari and Kr\"{u}mpelmann~\cite{Gari}.
This formalism incorporates $SU_{F}(3)$ symmetry relations and
Sakurai's universality hypothesis to describe the baryon
octet EM form factors, predominantly constrained by nucleon data.
Our treatment provided a good quantitative description of the data
using the vector meson-nucleon couplings,
$C_{\rho}(N) = 0.4$, $C_{\omega}(N) = 0.2$ and $C_{\phi}(N) = -0.1$
(see Ref.~\cite{WKL} where the value of $C_{\phi}(N)$ was optimized to
describe
$G_{E}^{n}$ data).
Here $C_{V}(N) = g_{VNN}/f_{V}$ is the ratio of the vector meson-nucleon
hadronic
coupling, $g_{VNN}$, to the vector meson-leptonic decay constant, $f_{V}$. 
Using
\begin{equation}
\Gamma_{V \rightarrow e^+ e^-} \;=\;
\frac{4 \pi \alpha_e^2}{3} \;  \frac{M_V}{f_V^2} 
\end{equation}
for the $\phi
\rightarrow e^+ e ^-$ decay yields, $f_{\phi} =
-13.1$, giving the $\phi N$  coupling $g_{\phi N N} = 1.3$. Recent, preliminary
$G_E^n$ measurements from
Jefferson Lab indicate a reduction at higher momentum compared to previous data,
which suggests an even
larger 
$\phi N$ coupling.   Although there
is  uncertainty in $g_{\phi N N}$,
which is governed by the currently unknown nucleon
strangeness content (as well as the small, but better known,
u and d quark content of the $\phi$),   the value $g_{\phi NN}=1.3$ 
accurately reproduces~\cite{TVCS}  both old, low, and new, high, $t$
$\phi$ photoproduction data.
The ratio of the $\phi N$ to the $\omega N$ coupling constant 
is $g_{\phi N N}/g_{\omega N N} = 0.37$, slightly
smaller but still consistent with Ref.~\cite{Ellis1}.

The $\gamma \pi \rightarrow \gamma_v$, $\gamma \eta \rightarrow \gamma_v$
and $\gamma G  \rightarrow \gamma_v$ transition form factors
are computed from VMD using the vector meson
propagators, $\Pi_{V=\rho,\omega,\phi}(q^2)$, with observed widths, $\Gamma_V$,
\begin{eqnarray}
\Pi_V(q^2) &=& - \frac{M_V^2}{q^2 - M_V^2 + i \sqrt{q^2} \; \Gamma_V}
\end{eqnarray}
and couplings determined directly from the
$\phi \rightarrow \gamma \pi$, $\phi \rightarrow \gamma \eta$,
$\omega \rightarrow \gamma \pi$, $\omega \rightarrow \gamma \eta$,
$\rho \rightarrow \gamma \pi$ and $\rho \rightarrow \gamma \eta$
decays
\begin{eqnarray}
\Gamma_{V \rightarrow X \gamma} \;=\;
\frac{\alpha_e}{3} \;  \; \frac{k^3_X}{M_0^2} \; \kappa_{X V
\gamma}^2 
\label{xVgamma}
\end{eqnarray}
for $X= \pi,\eta$ and $G$.
Again the mass, $M_0 = 1.0$ GeV, is an arbitrary scale  in the
$XV\gamma$
Lagrangian  and $k_X$ is the rest frame  3-momentum 
of the decay particles, given by a relation identical to Eq. (\ref{kv}).
VMD then yields
\begin{eqnarray}
\kappa_{\pi \gamma \gamma} \ F_{_{\gamma \pi \rightarrow \gamma_v}}(q^2) &=& 
\sum_{V=\rho, \omega, \phi} 
C_{\pi V \gamma} \; \Pi_{V}(q^2) \label{pi2g} \\
\kappa_{\eta \gamma \gamma} \ F_{_{\gamma \eta \rightarrow \gamma_v}}(q^2) &=& 
\sum_{V=\rho, \omega, \phi} 
C_{\eta V \gamma} \; \Pi_{V}(q^2) \label{eta2g} \\
\kappa_{G \gamma \gamma} \ F_{_{\gamma G \rightarrow \gamma_v}}(q^2) &=& 
\sum_{V=\rho, \omega, \phi} 
C_{G V \gamma} \; \Pi_{V}(q^2) \label{G2g} \ ,
\end{eqnarray}
where the dimensionless $C$-coefficients  
are ratios of transition moments and decay constants
\begin{eqnarray}
C_{\pi V \gamma} &=& \frac{\kappa_{\pi V \gamma}}{f_V} \\
C_{\eta V \gamma} &=& \frac{\kappa_{\eta V \gamma}}{f_V} \\
C_{G V \gamma} &=& \frac{\kappa_{G V \gamma}}{f_V} \ .
\end{eqnarray}
The pomeron/glueball radiative and transition 
decay constants are not directly known, however they
have been indirectly extracted from  $\phi$ photoproduction 
data and the VMD/universality relation~\cite{RAWphi,TVCS}
\begin{eqnarray}
\kappa_{G V \gamma} &=& g_{G V V'} \;
\left[ \frac{1}{f_\rho} + \frac{1}{f_\omega} + \frac{1}{f_\phi} \right]
= 0.62 \ .
\end{eqnarray}
Using the most recently measured  vector
meson radiative and leptonic
decay widths~\cite{PDG}, we obtain the VMD coupling constants
summarized in Table III.
% Table III
%%%%%%%%%%%%%%%%%%%%%%%%%%%%%%%%%%%%%%%%%%%%%%%%%%%%%%%%%%%%%%%%%%%%%%%
\begin{table}
\caption{VMD couplings  from  measured decays~\cite{PDG}.
The flavor independent glueball couplings  
are from Refs.~\cite{RAWphi,TVCS}.}
\begin{center}
\begin{tabular}{|c||c|c|c|c|}
\hline
$V$ & $\kappa_{\pi V \gamma}$ & $\kappa_{\eta V \gamma}$ &
$\kappa_{G V \gamma}$ & $f_{_V}$  \\
\hline
$\rho$ & 0.901 & 1.470  & 0.62 & 5.0 \\
$\omega$ & 2.324 & 0.532 & 0.62 & 17.1 \\
$\phi$ & 0.138 & 0.715 & 0.62 & -13.1 \\
\hline
\end{tabular}
\end{center}
\end{table}

The  pseudoscalar
$\pi \rightarrow \gamma \gamma$ and $\eta \rightarrow \gamma
\gamma$ radiative  decay widths
\begin{eqnarray}
\Gamma_{X\rightarrow \gamma \gamma} \;=\;
\frac{\pi \alpha_e^2}{4} \; \frac{M_X^3}{M_0^2} \; \kappa^2_{X
\gamma \gamma}
\label{x2gamma} \ ,
\end{eqnarray}
provide a VMD consistency check for the
$\pi$ and $\eta$ transition form factors, Eqs. (\ref{pi2g}) and (\ref{eta2g}), due to the
normalization conditions ($F_{\gamma \pi \rightarrow \gamma_v}(0) = 1$, etc.)
\begin{eqnarray}
\kappa_{\pi \gamma \gamma} &=& \frac{\kappa_{\pi \rho \gamma}}{f_{\rho}}
\;+\;   \frac{\kappa_{\pi \omega \gamma}}{f_{\omega}}
 \;+\;   \frac{\kappa_{\pi \phi \gamma}}{f_{\phi}} \label{pi2gamma}\\
\kappa_{\eta \gamma \gamma} &=& \frac{\kappa_{\eta \rho \gamma}}{f_{\rho}}
\;+\;   \frac{\kappa_{\eta \omega \gamma}}{f_{\omega}}
 \;+\;   \frac{\kappa_{\eta \phi \gamma}}{f_{\phi}}  \label{eta2gamma} \ . 
 \end{eqnarray}
Using recent data~\cite{PDG} we obtain excellent
agreement between experiment and the VMD predictions 
\begin{eqnarray}
{\rm experiment \; [Eq. \, (\ref{x2gamma})]:} &&  \; \kappa_{\pi \gamma \gamma} \;=\;
0.27  \;\;\;\;\;
		    \kappa_{\eta \gamma \gamma} \;=\; 0.26 \nonumber \\
{\rm VMD \; [Eqs. \, (\ref{pi2gamma},\ref{eta2gamma})] :} &&  \; \kappa_{\pi \gamma
\gamma} \;=\; 0.30
\;\;\;\;\;
	   \kappa_{\eta \gamma \gamma} \;=\; 0.27 \ . \nonumber
\end{eqnarray}

The glueball hadronic widths, using
\begin{eqnarray}
\Gamma_{G \rightarrow V V} \;=\;
\frac{g_{GVV}^2}{4 \pi} \; \frac{k^3_V}{M_0^2} \ ,
\label{x2V}  
\end{eqnarray}
are listed in
Table IV.  The notation $N/A$ indicates not allowed by isospin conservation,
while $N/P$
% Table IV
%%%%%%%%%%%%%%%%%%%%%%%%%%%%%%%%%%%%%%%%%%%%%%%%%%%%%%%%%%%%%%%%%%%%%%%
\begin{table}[b]
\caption{VMD glueball hadronic decay widths in MeV.}
\begin{center}
\begin{tabular}{|c||c|c|c|}
\hline
$\Gamma_{G \rightarrow V V'}$ & $\rho$ & $\omega$ & $\phi$   \\
\hline
\hline
$\rho$ & 44.4 & $\; N/A\;$  & $\;N/A\;$  \\
\hline
$\omega$ & $N/A$ & 34.6 & $N/P$ \\
\hline
$\phi$ & $N/A$ & $N/P$ & $N/P$  \\
\hline
\end{tabular}
\end{center}
\end{table}
represents insufficient  phase space.
It is noteworthy  that the $G \rightarrow $ $ \rho \rho$ and $\omega \omega$
widths of 44.4 and 34.6 MeV reasonably compare  to predictions from an independent
glueball mixing and decay analysis~\cite{bp} which predicts 46 and 12 MeV, respectively. 
The electromagnetic widths, using 
%\noindent
Eq. (\ref{xVgamma}), are
presented in Table V. The  vector meson photon decay widths  are also shown
for comparison.   
% Table IV
%%%%%%%%%%%%%%%%%%%%%%%%%%%%%%%%%%%%%%%%%%%%%%%%%%%%%%%%%%%%%%%%%%%%%%%
\begin{table}[h]
\caption{VMD  glueball electromagnetic decays in eV.}
\begin{center}
\begin{tabular}{|c||c|c|c|}
\hline
$V \rightarrow$ & $\rho$ & $\omega$ & $\phi$   \\
\hline
\hline
$\Gamma_{G \rightarrow V \gamma}$   & $\;866\;$ & $\;844\;$  &
$\;454\;$  \\
\hline
\hline
$\Gamma_{V \rightarrow \pi \gamma}$  & 102 & 717 & 6  \\
$\Gamma_{V \rightarrow \eta \gamma}$   & 36 & 6 & 59  \\
\hline
\end{tabular}
\end{center}
\end{table}

We can also calculate the glueball 2-photon
radiative decay width by first evaluating Eq. (\ref{G2g}) for $q^2 = 0$,
\begin{eqnarray}
\kappa_{G \gamma \gamma} &=& \kappa_{G V \gamma} \
\left[ \frac{1}{f_\rho} + \frac{1}{f_\omega} + \frac{1}{f_\phi} \right]
\ = 0.11
\end{eqnarray}
where again universality is invoked for all
$V$.  The VMD
prediction from Eq. (\ref{x2gamma}) for the $G \rightarrow \gamma \gamma$ decay width is then,
$\Gamma_{G \rightarrow \gamma \gamma} = 2.6$ eV, which is comparable to the meson decay widths
$\Gamma_{\pi \rightarrow \gamma \gamma}= 7.74$
eV and
$\Gamma_{\eta \rightarrow \gamma \gamma}= 0.46$ eV.

Combining the above values yields a total hadronic $VV$ scalar glueball width of 79 MeV
which of course only represents a lower bound for the full width.  Indeed there are
several other decay channels involving lighter mesons ($\pi$, $\eta$, $K$, $a_1$) which
will compete and Ref.~\cite{bp} estimates the total width could be up to 250
MeV.   The actual width is  between these two limits and probably closer to
the observed widths for the $f_0(1500)$ and $f_0(1710)$ glueball candidates which are
109 and 125 MeV, respectively.

Finally,  glueball electroproduction via intermediate 
virtual photons will require form factors for the transition $\gamma_v V \rightarrow G$. 
Again, simple application of VMD yields the appropriate $\gamma_v V \rightarrow G$ transition 
form factors
\begin{eqnarray}
\kappa_{G V \gamma} \ F_{_{\gamma V \rightarrow G}}(q^2) \;=\; 
\sum_{V' = \rho, \omega, \phi} \;  \frac{g_{GVV'}}{f_{V'}} \; \Pi_{V'}(q^2) \;.
\end{eqnarray}

\vspace{-.4cm}

%%%%%%%%%%%%%%%%%%%%%%%%%%%%%%%%%%%%%%%%%%%%%%%%%%%%%%%%%%%%%%%%%%%%%
%%%%%%%%%%%%%%%%%%%%%%%%%%%%%%%%%%%%%%%%%%%%%%%%%%%%%%%%%%%%%%%%%%%%%

%\newpage

\section{Cross Section Predictions}

This section summarizes the key cross section findings and presents results
for a variety of kinematics.  Figure 1 displays the exclusive 
photoproduction  cross section versus the $VV$ invariant mass. 
The  three solid curves, corresponding to different glueball widths, represent production and
decay to the 
$\rho
\rho$ or
$\omega
\omega$ final states.  Because of universality ($g_{G \rho \rho} = g_{G \omega \omega}$) and the
near degeneracy of the $\rho$ and $\omega$ masses, the $\rho \rho$ and $\omega \omega$
production cross sections are essentially equal so one curve represents both channels.
The two short dashed curves are for  $\phi \phi$ production which has a
higher threshold.  

The long dashed curve is the non-Regge prediction using $\Gamma_G = $ 79 MeV.
This is the result using the gluonic propagator without the Regge factor
and, as discussed previously, can be regarded as production mediated by more conventional
hadron (glueball) formation.  Notice that it is about a factor of 4 larger than
the Regge (pomeron) prediction using the same width (top solid line).  A similar 
increase occurs for the two other widths (not shown). 

%\newpage

%%%%%%%%%%%%%%%%%%%%%%%%%%%%%%%%%%%%%%%%%%%%%%%%%%
\begin{figure} 
\epsfxsize = 3.5in
\epsfysize = 4in
%\epsffile{dsm.ang0e5.8.ps}
\epsffile{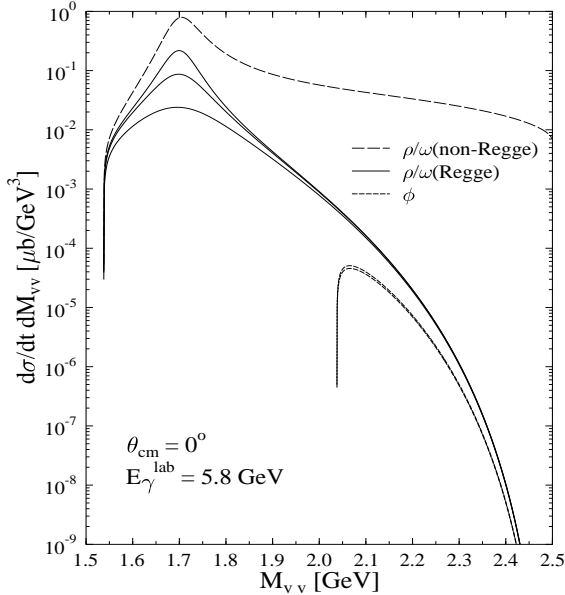}
\caption{Cross sections for $p(\gamma, G \rightarrow VV)p$ vs. the invariant $VV$ mass.
Solid lines represent the 
$\rho\rho$ (or $\omega \omega$) cross sections for  different glueball widths.  The
short dashed curves represent 
$\phi \phi$ photoproduction. The long dashed curve omits the Regge energy dependence.}
\end{figure}

%%%%%%%%%%%%%%%%%%%%%%%%%%%%%%%%%%%%%%%%%%%%%%%%%%
\begin{figure} 
\epsfxsize = 3.5in
\epsfysize = 4in
%\epsffile{dsm.ang0e5.8terms.ps}
\epsffile{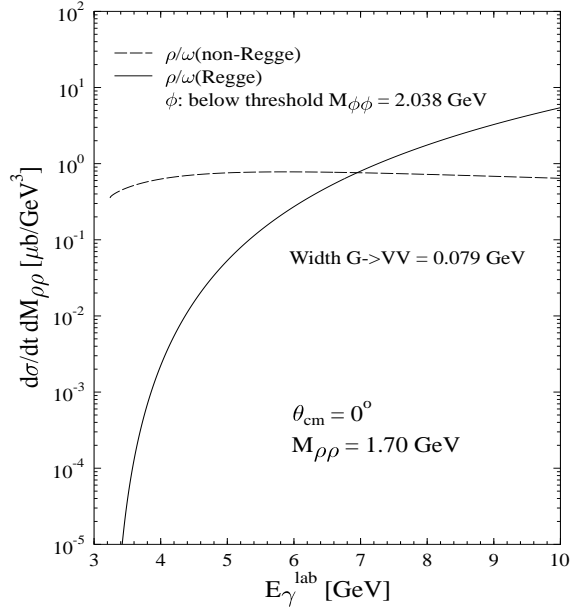}
\caption{Comparison of Regge (pomeron, solid line) and non-Regge (glueball, dashed line)
mediated production vs. lab energy.}
\end{figure}

%%%%%%%%%%%%%%%%%%%%%%%%%%%%%%%%%%%%%%%%%%%%%%%%%%
\begin{figure} 
\epsfxsize = 3.5in
\epsfysize = 4in
%\epsffile{dsm.ang0e5.8terms.ps}
\epsffile{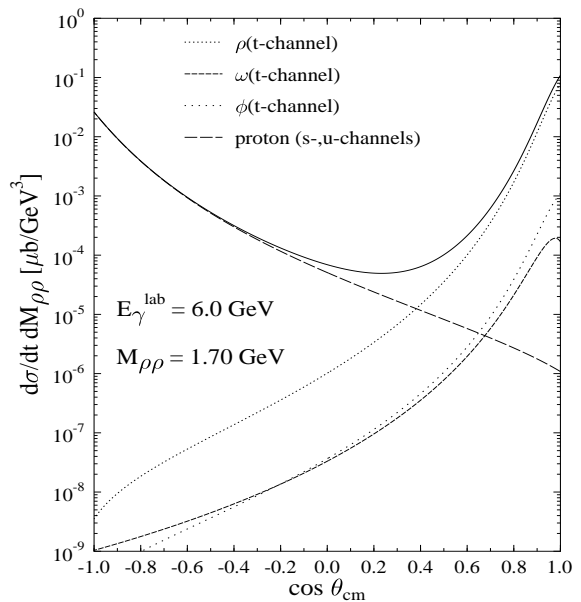}
\caption{Relative $s$, $t$ and $u$-channel contributions.}
\end{figure}

%%%%%%%%%%%%%%%%%%%%%%%%%%%%%%%%%%%%%%%%%%%%%%%%%%
%\begin{figure} 
%\epsfxsize = 3.5in
%\epsfysize = 4in
%\epsffile{dsm.ang90e5.8.ps}
%\epsffile{Fig5.ps}
%\caption{Same as Fig. 1 for $\theta_{cm}$ = $90^o$.}
%\end{figure}

%\newpage

%\newpage

To document sensitivity to the uncertain glueball width, three 
values are displayed.  The upper solid
curve depicting the distinct resonant glueball structure assumes that vector meson decay
saturates the entire glueball width and uses the
value $\Gamma_{G \rightarrow V V} =
\Gamma_{G \rightarrow \rho \rho} + 
\Gamma_{G \rightarrow \omega \omega} =$ 79 MeV from Table IV. The 
lower curve corresponds to a   width of 238 MeV which is taken as an upper bound and
is also the numerical
width necessary to completely suppress the glueball cross section enhancement.
The middle curve uses $\Gamma_G = 125$ MeV which,  
as discussed in  section IV, is probably closer to the physical glueball value.
Hence, if the actual width is roughly of order
100 MeV, a clear glueball enhancement can emerge.

%\newpage

The two dashed $\phi \phi$ production
curves, corresponding to the minimum and maximum glueball widths, are
essentially the same since the
$\phi
\phi$ threshold is well above the on-shell glueball mass ($M_G = 1.7 $ GeV) and only a small
effect is present from the off-shell gluonic propagator.
Consequently $\phi \phi$ production is predicted to be devoid of a light scalar glueball
enhancement.    This is
also why the  $\rho \rho$ (or $\omega \omega$) curves converge at higher invariant
$VV$ mass, a region of interests for effects from a
tensor,
$J^{PC} = 2^{++}$, glueball that is expected to have mass near  2 GeV.  Related,
Ref.~\cite{PDG} lists several $f_2$, possible glueball, states above 2 GeV with observed $\phi
\phi$ decays.

%\newpage

The cross section kinematics reflect the capability of
the  envsioned Hall D facility at Jefferson Lab.  
Depending on choice of gluonic propagator, 
there is  
sensitivity to the incident photon energy as indicated in Fig. 2.  While the energy behavior of
the non-Regge prediction (dashed line) is relatively flat, the Regge calculation (solid curve)
increases with higher beam energy.  There is
little energy or
$s$ dependence in the non-Regge calculation since the cross section is at forward angles
($t$-channel dominated).   
Because the Regge formulation is more phenomenologically based, it is our preferred
prediction.  However, it would be interesting to confront both results with multiple
vector meson production data at low and high energies.  Interestingly, the two formulations
yield similar cross sections (and measurable production rates) near 7 GeV   which is
a frequently cited  Jlab upgraded photon energy.  Hence it should be feasible to confirm
the gluonic enhancement predicted in Fig. 1.

\newpage

There is significant sensitivity to the production final state angle,
$\theta_{cm}$,
or momentum transfer.  This is reflected in Fig. 3 which details a falling, then rising
cross section with increasing angle.  Note the minimum cross section occurs for
$\theta_{cm}$ = $90^{o}$.  This prediction is for $E^{lab}_{\gamma} =$ 6.0 GeV
and corresponds to the intermediate, more representative glueball width.
As expected for small angles (low $t$)  $t$-channel vector
meson exchange dominates with $\rho$ exchange (dense dot curve) being the most important.  At
larger angles (higher $t$) $s$ and $u$-channel amplitudes emerge (dashed curve) corresponding
to production from glueball-proton coupling.  Because both $s$ and $q'$ are fixed, the
Regge and non-Regge (not shown) results have identical $t$-channel relative contributions and
variations.

%\newpage 
%\vspace{.5cm}

It is important to relate these predictions to the expected background $VV$ production
from non-glueball mediated processes.  Unfortunately, to our knowledge there are no 
specific model calculations in this energy region.  However, other double
meson photoproduction  calculations exist and 
Refs.~\cite{as,bls} predict cross sections comparable in magnitude to those calculated
here. Further, Ref.~\cite{as} investigates possible exotic, $J^{PC} =1^{-+}$, meson excitation
in $\gamma p \rightarrow \rho^0 \pi^+ n$ and predicts a similar resonance profile to
our glueball production result. 

%\newpage

Finally, we note a novel detection signature predicted by this analysis.
Because of the dominant $\rho \rightarrow \pi \pi$, $\omega \rightarrow \pi \pi \pi$
and $\phi \rightarrow K K$ decays, the presence of a glueball excitation
should be correlated with a 4 and 6 $\pi$ decay around 1.7 GeV in the invariant
$\rho \rho$ and $\omega \omega$  mass spectra, respectively. 
Further, and depending on the proximity of the glueball mass to the $\omega \phi$ threshold,
there also may be a $\pi^+
\pi^-
\pi^0 K^+ K^-$ decay near or above 1.8 GeV in the $\omega \phi$ spectrum.  The latter may be a
unique signature as there are no hadrons  listed with this decay.  This would be an ideal
experiment for the envisioned Hall D large acceptance spectrometer.

%%%%%%%%%%%%%%%%%%%%%%%%%%%%%%%%%%%%%%%%%%%%%%%%%%
%%%%%%%%%%%%%%%%%%%%%%%%%%%%%%%%%%%%%%%%%%%%%%%%%%%%%%%%%%%%%%%%%%%%%%%%%%%

%\newpage

\section{Conclusions}

This work combines the time-honored tools of Quantum Hadrodynamics, vector meson
dominance and Regge theory with the pomeron-glueball connection hypothesis to
predict glueball production and decay processes.  Using a minimal set of parameters
independently determined from recent hadronic and electromagnetic analyses,
a measurable cross section is predicted for $p(\gamma, G \rightarrow V V)p$.
Most significant is a possible scalar glueball enhancement near 1.7 GeV 
in the $\rho \rho$ and $\omega \omega$ invariant mass spectra.  This resonant cross
section structure is sensitive to the total glueball width and should be discernable if the width
is of order 100 MeV.  If the actual width is significantly larger, say
greater than 200 MeV, it may still be possible to detect a scalar glueball via
decay to either 4 or 6 pions having an invariant mass near 1.7 GeV.  Even more novel
would be a correlated $\pi^+ \pi^- \pi^0 K^+ K^-$ observation in this same mass region,
which would be a unique decay signature from a scalar hadron.  Such measurements
would be ideal for the envsioned JLab energy upgrade and new Hall D 
wide acceptance spectrometer.

Future work will address other scalar glueball decay channels. Related,
tensor glueball production and decay will be investigated which will be
especially interesting since the $2^{++}$ glueball mass is expected to
be above the clear signature $\phi \phi$ decay threshold.

\begin{acknowledgments}
This work was supported by the Department of Energy under grant
DE-FG02-97ER41048.
\end{acknowledgments}

%\newpage

%%%%%%%%%%%%%%%%%%%%%%%%%%%%%%%%%%%%%%%%%%%%%%%%%%%%%%%%%%%%%%%%%%%%%%%%%
%\noindent {\Large \bf{References}} \\

\end{document}